\documentclass[fleqn,usenatbib]{mnras}

\usepackage{newtxtext,newtxmath}

\usepackage[T1]{fontenc}

\DeclareRobustCommand{\VAN}[3]{#2}
\let\VANthebibliography\thebibliography
\def\thebibliography{\DeclareRobustCommand{\VAN}[3]{##3}\VANthebibliography}

\usepackage{graphicx}	
\usepackage{amsmath}	

\newcommand{\UV}{\mbox{$\mathrm{UV_{275}}$}}
\newcommand{\U}{\mbox{$\mathrm{U_{336}}$}}
\newcommand{\B}{\mbox{$\mathrm{B_{438}}$}}
\newcommand{\V}{\mbox{$\mathrm{V_{606}}$}}

\newcommand{\I}{\mbox{$\mathrm{I_{814}}$}}

\title[Millisecond pulsars of M13]{Chandra and HST Studies of Six Millisecond Pulsars in the Globular Cluster M13}

\author[Zhao, Zhao \& Heinke]{
Jiaqi Zhao,$^{1}$\thanks{E-mail: jzhao11@ualberta.ca}
Yue Zhao,$^{1}$
Craig O. Heinke$^{1}$
\\
$^{1}$Department of Physics, University of Alberta, CCIS 4-183, Edmonton, AB T6G 2E1, Canada
}

\date{Accepted XXX. Received YYY; in original form ZZZ}

\pubyear{2020}

\begin{document}
\label{firstpage}
\pagerange{\pageref{firstpage}--\pageref{lastpage}}
\maketitle

\begin{abstract}
We analyse 
55 ks of {\it Chandra} X-ray observations of the Galactic globular cluster M13. 
Using the latest radio timing positions of six known millisecond pulsars (MSPs) in M13 from Wang et al. (2020), we detect confident X-ray counterparts to five of the six MSPs at X-ray luminosities of $L_X$(0.3-8 keV)$\sim 3 \times 10^{30} - 10^{31}~{\rm erg~s^{-1}}$, including the newly discovered PSR J$1641+3627$F. There are limited X-ray counts at the position of PSR J$1641+3627$A, for which we obtain an upper limit 
$L_X<1.3 \times 10^{30}~{\rm erg~s^{-1}}$. We analyse X-ray spectra of all six MSPs, which are well-described by either a single blackbody or a single power-law model. We also incorporate optical/UV imaging observations from the {\it Hubble Space Telescope (HST)} and find optical counterparts to PSR J$1641+3627$D and J$1641+3627$F. 
Our colour-magnitude diagrams indicate the latter contains a white dwarf, consistent with 
the properties suggested 
by radio timing observations. The counterpart to J$1641+3627$D is only visible in the V band; however, we argue that the companion to J$1641+3627$D is also a white dwarf, since 
we see a blackbody-like X-ray spectrum, while MSPs with nondegenerate companions generally show non-thermal X-rays from shocks between the pulsar and companion winds. 
Our work increases the sample of known X-ray and optical counterparts of MSPs in globular clusters.
\end{abstract}




\section{Introduction}
\label{sec: Introduction}

Radio millisecond pulsars (MSPs), also known as rotation-powered MSPs, are fast-spinning pulsars (spin periods $P \lesssim 30$ ms) with low spin-down rates ($\Dot{P} \sim 10^{-21} - 10^{-19}$), implying large characteristic ages $\tau \equiv P/(2 \Dot{P}) \gtrsim 1$ Gyr, and relatively low 
magnetic field strengths $B_p \propto (P \Dot{P})^{1/2} \sim 10^8 - 10^{10}$ G. Low-mass X-ray binaries (LMXBs) are conventionally considered the progenitors of MSPs, where a neutron star (NS) is spun up by accreting material from its companion star until it has a rotational period of a few milliseconds \citep{Alpar82,Bhattacharya91}. The high stellar densities of globular clusters (GCs) create numerous LMXBs, which then produce MSPs, and hence GCs provide ideal places to observe them in large numbers \citep{Camilo05}. To date, the total number of pulsars found in 30 GCs is 157, and more than 90\% of them are MSPs\footnote{For an up-to-date catalog of pulsars in GCs, see \url{http://www.naic.edu/~pfreire/GCpsr.html}}. 

MSPs are mostly found in binary systems, which is consistent with the evolution of MSPs via LMXBs. The so-called ``spider'' MSP binaries represent a distinct group of MSP binary systems with low-mass nondegenerate companion stars. Specifically, spider binaries are classified into two groups based on the companion masses: black widows with companion masses $M_c \sim 0.02~{\rm M}_{\sun}$, and redbacks with companion masses $M_c \sim 0.2~{\rm M}_{\sun}$ \citep{Roberts11}. Alternatively, 
 MSPs may be coupled with compact objects, such as helium-core white dwarfs (WDs), which are the most common companions to MSPs in GCs \citep[e.g.][]{Camilo05}. 
 MSPs coupled with another neutron star or even a detected radio pulsar have been discovered in a few systems, like PSR J$0737-3039$ 
 \citep{Burgay03}. 

Radio MSPs are also faint X-ray emitters, with typical luminosities of $L_X \sim 10^{30} - 10^{31}~{\rm erg~s^{-1}}$. For a few relatively young and energetic MSPs, like PSR B1821$-$24, the X-ray luminosities can reach up to $\sim 10^{33}~{\rm erg~s^{-1}}$ \citep[e.g.][]{Bogdanov11}. The X-rays produced by MSPs can be characterized based on their spectral properties, namely thermal (blackbody-like spectra) or non-thermal (power-law spectra) emission \citep{Becker99,Bogdanov18}. Thermal X-ray emission is believed to be generated from the hot surface of the NS,  specifically from the hot spots near the magnetic polar caps, heated by the return flow of relativistic particles from the pulsar magnetosphere \citep[e.g.][]{Harding02}. Non-thermal X-ray emission can be further categorized into two sub-groups, i.e. pulsed and non-pulsed non-thermal emission. Pulsed non-thermal X-rays are observed with narrow X-ray pulsations, implying highly beamed X-ray radiation, which is most likely produced in the pulsar magnetosphere 
\citep{Verbunt96,Saito97,Takahashi01}.
Therefore, only very energetic MSPs with relatively strong 
spin-down luminosities could emit such X-ray radiation \citep[e.g.][]{Possenti2002}. Non-pulsed non-thermal X-ray emission is commonly detected from spider pulsar systems, where the relativistic pulsar wind may collide with the material from its companion star, creating an  intra-binary shock and emitting non-pulsed, non-thermal X-rays \citep[e.g.][]{Arons93,Stappers03,Bogdanov05,Gentile14,Roberts15}. 

The globular cluster M13 (NGC 6205) is located in the constellation of Hercules, 
with a low foreground reddening of  $E(B-V)=0.02$ \citep[][2010 edition]{Harris96}. 
The distance is slightly uncertain, with a range of reported values (mostly isochrone fitting to the colour-magnitude diagram, but also using RR Lyrae variables and the tip of the red giant branch) from 7.1$\pm0.1$ kpc \citep[][(2010 revision)]{Deras19,Harris96} to 7.9$\pm0.5$ kpc \citep{Barker18,Sandquist10}. We use the distance of 7.4$\pm0.2$ kpc from the recent comprehensive study of \citet{Gontcharov20} in this work. Should the true distance lie at 7.1 or 7.9 kpc, then values of luminosities in this paper would change by up to 10\%.

To date, six MSPs have been found in M13 by several radio surveys \citep{Kulkarni91,Anderson93,Hessels07,Wang20}. PSRs J$1641+3627$A (hereafter MSP A) and J$1641+3627$C (hereafter MSP C) are isolated, while the other four MSPs are in binary systems. Specifically, PSR J$1641+3627$E (MSP E) is found in an eclipsing black widow system, with a minimum companion mass of $\sim 0.019~{\rm M}_\odot$ \citep{Wang20}. The nature of the companions for PSRs J$1641+3627$B (MSP B), J$1641+3627$D (MSP D), and J$1641+3627$F (MSP F) is not clear so far. New, precise radio timing solutions for the six MSPs in M13 were reported recently by \citet{Wang20} using  Five-hundred-meter Aperture Spherical radio Telescope (FAST) observations. 

In this paper, we present X-ray spectral analyses of the six MSPs in M13 using archival {\it Chandra} observations. We also investigate the optical counterparts to those MSPs in binary systems, based on observations from the {\it Hubble Space Telescope} ({\it HST}). This work is organized as follows. In section~\ref{sec: Observation and data reduction}, we describe the observations and data reduction procedures. In section~\ref{sec: Data analysis and results}, we present the X-ray spectral fitting results for the six MSPs, and the search for counterparts to the MSPs in optical/UV bands. We discuss the X-ray spectral properties and the nature of the companion stars in section~\ref{sec: Discussion}. Finally, we draw conclusions in section~\ref{sec: Conclusion}.

\section{Observation and data reduction}
\label{sec: Observation and data reduction}
\subsection{Chandra observations}

The X-ray data used in this work consists of two {\it Chandra X-ray Observatory} observations of M13 in 2006, 
with a total exposure time of 54.69 kiloseconds (see Table~\ref{tab:obs}). 
For both observations, the core of M13 was positioned on the back-illuminated ACIS-S3 chip and configured in FAINT mode. 

\begin{table}
	\centering
	\caption{\textit{Chandra} Observations of M13}
	\label{tab:obs}
	\begin{tabular}{lccr} 
		\hline
		Telescope/ & Date of & Observation & Exposure\\
		Instrument & Observation & ID & Time (ks) \\
		\hline
		$\textit{Chandra}$/ACIS-S & 2006 Mar 09 & 7290 & 27.9 \\
        $\textit{Chandra}$/ACIS-S & 2006 Mar 11 & 5436 & 26.8 \\
		\hline
	\end{tabular}
\end{table}

The data reduction and analysis were performed using {\sc ciao}\footnote{Chandra Interactive Analysis of Observations, available at \url{https://cxc.harvard.edu/ciao/}} (version 4.12, CALDB 4.9.1, \citealt{Fruscione06}). We first reprocessed the data with the \texttt{chandra\_repro} script to generate new level 2 event files of the observations, applying the newest calibration updates and bad pixel files. We filtered the data to the energy range 0.5$-$7 keV, where 
the X-ray emission from MSPs may dominate over the instrumental background.
No background flares were seen in the \textit{Chandra} data.

\begin{figure*}
    \centering
    \includegraphics[width=\textwidth]{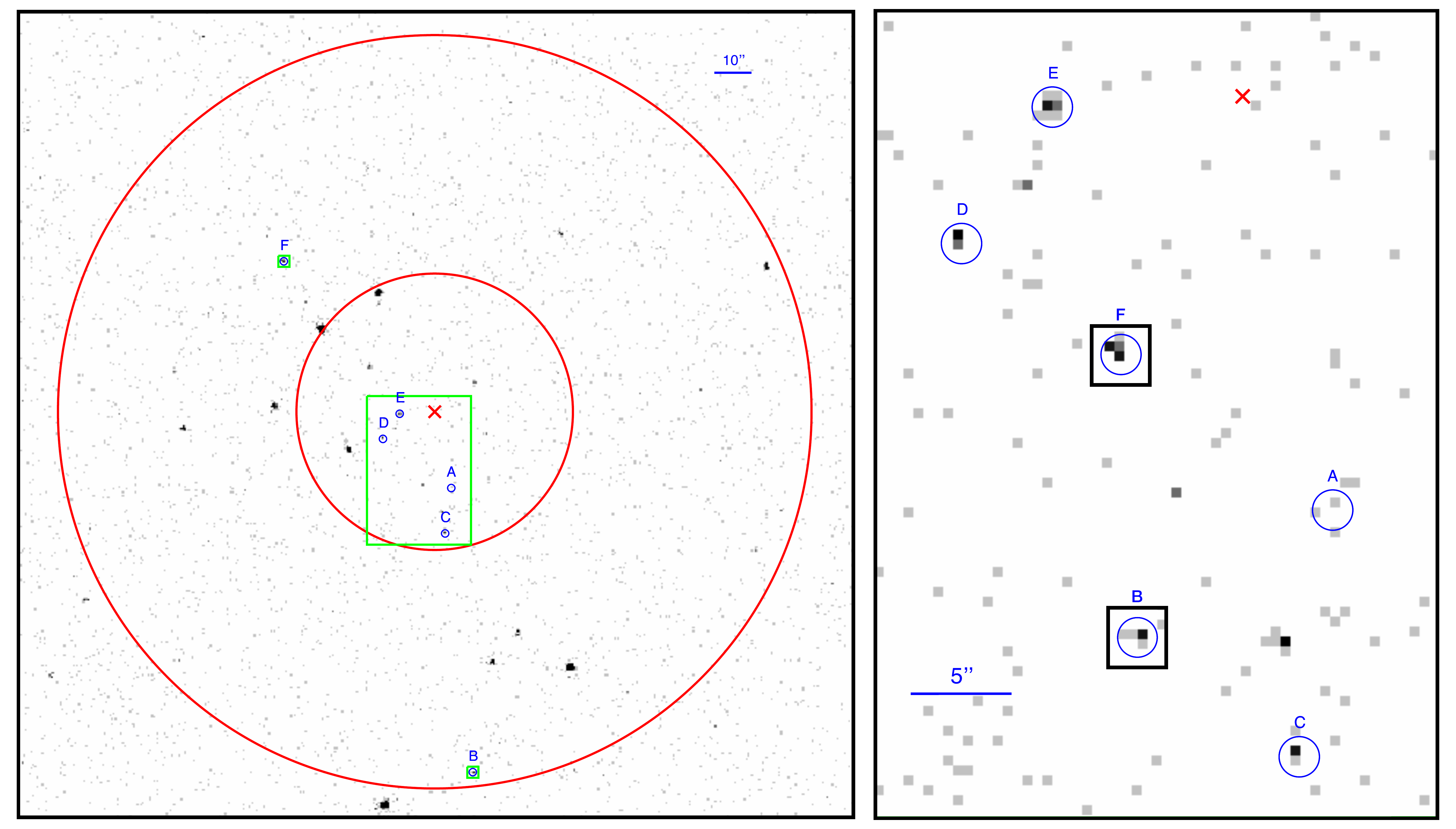}
    \caption{Left: merged 0.5$-$7 keV Chandra X-ray image of M13, with the positions of the 6 known MSPs labeled (blue circles with 1$\arcsec$ radii and letters). The center of M13 is marked with a red cross. The smaller red circle shows the 0\farcm62 core radius of M13, while the larger one shows the 1\farcm69 half-light radius of M13 \citep[2010 edition]{Harris96}. The green boxes surrounding the MSPs are detailed in the right figure. Right: the $28\arcsec \times 40\arcsec$ core region of M13 which includes MSPs A, C, D, and E (the largest green box in the left figure). MSPs B and F are inset at source-free places with the same scale. The X-ray emission from all MSPs but A are clearly visible. North is up, and east is to the left. Brighter X-ray sources visible in the left image include cataclysmic variables and a quiescent X-ray binary \citep[see][]{Servillat11,Shaw18}. }
    \label{fig:merged}
\end{figure*}

We created a co-added image of M13 by merging the event files from the two observations using \texttt{reproject\_obs} script. 
In Figure~\ref{fig:merged}, the positions of the six MSPs are marked by blue circles with 1$\arcsec$ radii, centered on the precise radio pulsar timing positions \citep{Wang20}. Other brighter X-ray sources are also visible in this image, including a quiescent low-mass X-ray binary \citep{Shaw18} and cataclysmic variables \citep[e.g.][]{Servillat11}. We formally detected the X-ray counterparts to most of the MSPs by applying the {\sc ciao} tool \texttt{wavdetect}, a Mexican-Hat Wavelet source detection tool \citep{Freeman02}. 
We specified the wavelet scales (a list of radii in pixels) of 1.0, 1.4, and 2.0, and a significance threshold for source detection of $10^{-4}$ (false sources per pixel). Consequently, five X-ray counterparts were detected, all but MSP A, with positions  consistent with the radio positions. 

To analyse the X-ray spectra of the MSPs, we extracted the emission from the circular regions with a radius of 1$\arcsec$ centered on the radio positions in energy band 0.5$-$7 keV (Figure~\ref{fig:merged}) for each MSP, using the \texttt{specextract} script. The extraction process was performed separately for each observation, and then we used the \texttt{combine\_spectra} script to co-add the spectra correspondingly for each pulsar to obtain the combined spectra for spectral analysis. The background was taken from source-free annular regions around the MSPs. 

\subsection{Optical observations}
We use imaging data taken by the Wide Field Camera 3 (WFC3; GO-12605)  and Advanced Camera for Surveys (ACS; GO-10775) on board the {\it HST}. GO-12605 (PI: Piotto) contains exposures in two UV filters, F275W ($\UV$) and F336W ($\U$), along with an exposure in F438W ($\B$); while GO-10775 (PI: Sarajedini) is comprised of exposures in $\V$ (F606W) and $\I$ (F814W). For all filters, we retrieved the FLC data products from the Mikulsky Archive for Space Telescope (MAST)\footnote{\url{https://archive.stsci.edu/hst/search.php}}; these are images that have been pipe-lined, flat-fielded, and have charge transfer efficiency trails removed. Detailed information on these observations is summarised in Table \ref{tab:hst_obs}.

To search for faint potential  counterparts to the MSPs, we use the {\sc drizzlepac} software (version 3.1.6)\footnote{\url{https://www.stsci.edu/scientific-community/software/drizzlepac}} to generate combined {\it HST} images. FLC files in each filter are first re-aligned by the {\tt TweakReg} tool to a reference image (chosen to be the longest FLC exposure) and then combined using the {\tt AstroDrizzle} tool. {\tt AstroDrizzle} corrects for geometric distortion, flags cosmic rays and small-scale detector defects, and combines images with user-defined re-sampling. We use $pixfrac=1.0$ and oversample the combined images by a factor of two, so the final images have half the original pixel scales ($0.02\arcsec$/pixel for WFC3, $0.025\arcsec$/pixel for ACS).

Starting from December 2019, MAST released updated absolute astrometry information for ACS and WFC3 data. Most FLC data products are now aligned to the Gaia DR2 catalogue, reducing the astrometric uncertainties to $\sim 10~\mathrm{mas}$\footnote{\url{https://archive.stsci.edu/contents/newsletters/may-2020/new-absolute-astrometry-for-some-hst-data-products}}. The observed M13 fields contain stars included in  Gaia DR2, so we use the default WCS information to set our absolute astrometry.

\begin{table*}
    \centering
    \begin{tabular}{ccccc}
    \hline
      GO        & Exposure (s) & Observation Start & Instrument & Filter\\
    \hline
      10775     & 567  & 2006-04-02 10:41 & ACS/WFC & F606W ($\V$) \\
      10775     & 567  & 2006-04-02 12:15 & ACS/WFC & F814W ($\I$) \\
    \hline
      12605     & 1281 & 2012-05-14 01:52 & WFC3/UVIS & F275W ($\UV$) \\
      12605     & 1281 & 2012-05-17 03:54 & WFC3/UVIS & F275W ($\UV$) \\
      12605     & 700  & 2012-05-14 02:24 & WFC3/UVIS & F336W ($\U$) \\
      12605     & 700  & 2012-05-17 04:48 & WFC3/UVIS & F336W ($\U$) \\
      12605     & 92   & 2012-05-14 01:49 & WFC3/UVIS & F438W ($\B$) \\
      12605     & 92   & 2012-05-17 03:26 & WFC3/UVIS & F438W ($\B$) \\
    \hline
    \end{tabular}
    \caption{HST observations used in this work.}
    \label{tab:hst_obs}
\end{table*}

\section{Data analysis and results}
\label{sec: Data analysis and results}

\subsection{X-ray spectral fits}
\label{subsec:spectral fits}

\begin{figure*}
    \centering
    \includegraphics[width=\textwidth]{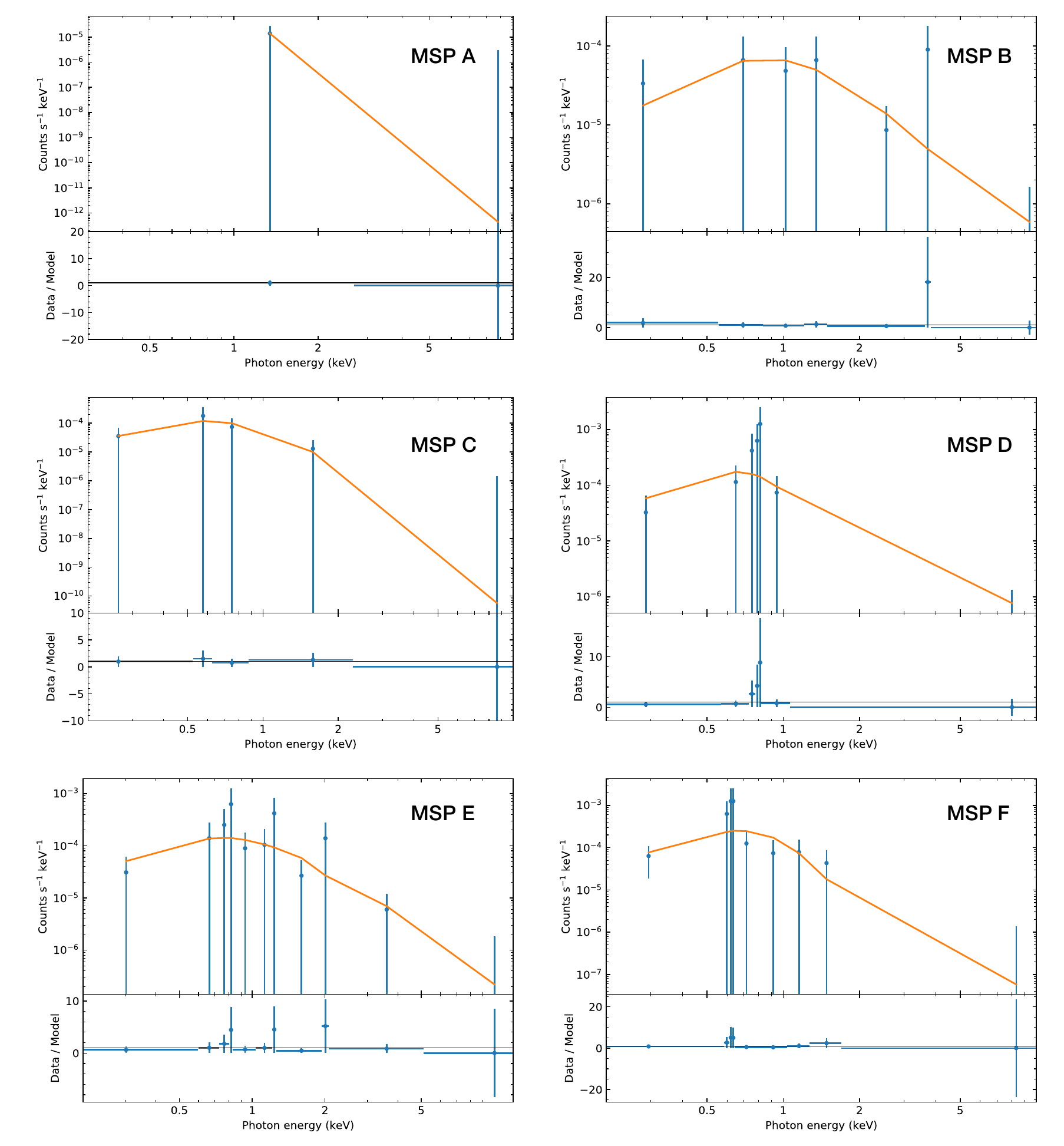}
    \caption{X-ray spectra and best fits for five MSPs in M13. The data are binned with 1 count/bin, and fitted using the WSTAT statistic. 
    }
    \label{fig:spectra}
\end{figure*}

We performed all spectral fits using  {\sc ciao}'s modeling and fitting application, Sherpa\footnote{Available at~\url{https://cxc.cfa.harvard.edu/sherpa/}}. X-ray emission from MSP A was only detected in the latter observation (Obs ID: 5436) with just two photons. We cannot  determine whether the two photons originated from MSP A, or are just  background emission. However, we fitted the spectrum of MSP A, and set the obtained fits as the upper limits, finding $L_X < 1.3\times10^{30}$ erg s$^{-1}$. The other five MSPs in M13 (MSPs B, C, D, E, and F) show faint and relatively soft X-ray emission  (Figure~\ref{fig:spectra}), with X-ray luminosities  $\sim 3\times10^{30}-10^{31}$ erg s$^{-1}$.  We adopted the WSTAT statistic\footnote{See \url{https://heasarc.gsfc.nasa.gov/xanadu/xspec/manual/XSappendixStatistics.html} for more details.}, a Poisson log-likelihood function including a  Poisson background, within Sherpa to fit  X-ray spectra with few photons. In addition, we grouped the data to include at least one photon in each bin due to the limited number of photons \citep{Humphrey2009}. 
For all six detected MSPs, we fitted the spectra by fixing the hydrogen column density ($N_{\rm H}$) to the cluster. 
We estimated $N_{\rm H}$ from the known reddening \citep[2010 edition]{Harris96} and an appropriate conversion factor \citep{Bahramian15} and obtained a value of $1.7 \times 10^{20}\ \text{cm}^{-2}$, given that interstellar extinction $E(B-V)$ generally gives the best predictions of $N_{\rm H}$ \citep{He2013}.

We considered 
spectral models of the X-ray emission from MSPs \citep{Bogdanov06} involving a blackbody (BB), a power-law (PL), and combinations of these. 
For the BB model, we used the \texttt{xsbbodyrad} model in Sherpa, and the free parameters were the effective temperature and the normalized radius. The PL model was fitted using \texttt{xspegpwrlw}, with the photon index and flux as free parameters. 
Figure~\ref{fig:spectra} shows the X-ray spectra and best fits of the six MSPs, and the best-fit models and parameters are given in Table~\ref{tab:fits}. We used the Q-value, which is a measure of what fraction of simulated spectra would have a larger value of the reduced statistic than the observed one, if the assumed model and the best-fit parameters are true, to indicate the goodness of the fits. The Q-values in Table~\ref{tab:fits} are above 0.05, indicating that these are reasonable fits and hence the Q-values themselves did not rule out any models. 

The spectra of all six MSPs in M13 are well described by either a pure BB model or a pure PL model.  (We tested BB+PL and BB+BB models, but these did not give better fits, so we only discuss simple one-component fits henceforth.) Although we cannot rule out either model from Q-values alone, other fitting parameters, like effective temperature and photon index, provide reasons to exclude models. For instance, if we fit the spectra of MSPs C, D, and F using a PL model, the obtained photon indices are nearly 4, which are empirically not observed from MSPs, given the typical photon indices of MSPs $\Gamma \sim 1.5$ \citep[e.g.][]{Zhang03,Bogdanov06,Bogdanov11}. Similarly, if we fit MSPs B and E using a single BB model, the fitted effective temperatures are too high while the fitted effective radii are too small for MSPs, compared to other MSPs in GCs \citep[e.g.][]{Bogdanov06,Forestell14}, and hence we can rule out the BB model for these two MSPs. Alternatively, the spectra of MSPs B and E are well described by a pure power-law model, with spectral photon indices $\Gamma = 1.8 \pm 0.7$ and $2.2 \pm 0.6$, respectively. Given that MSP E is a ``black widow'' pulsar, the bulk of its observed non-thermal X-rays are likely to be produced by interaction of the relativistic particle wind from the pulsar with matter lost from the companion. MSPs C, D, and F have X-ray spectra well fitted with a pure blackbody spectrum, implying no or little X-ray emission from a pulsar magnetosphere and/or intra-binary shock. Blackbody-like X-ray spectra are common from MSPs \citep[e.g.][]{Bogdanov06}, and likely originate from small hot spots at the magnetic poles heated by relativistic particles in the pulsar magnetosphere \citep[e.g.][]{Harding02}.

Particularly, since we only have two photons from MSP A, we need to fix one more parameter to obtain at least one degree of freedom. In order to determine the upper limit of the luminosity of MSP A, we fixed the effective temperature $T_{\rm eff}$ and photon index $\Gamma$ for the BB and PL models in the fitting processes, respectively. The value of the fixed $T_{\rm eff}$ was obtained by averaging the fitted $T_{\rm eff}$ of MSPs C, D, and F, giving a value of $T_{\rm eff} = 1.6 \times 10^6$ K, while the fixed $\Gamma$ was given by the mean value of photon indices of MSPs B and E, providing $\Gamma = 2.0$ (see Table~\ref{tab:fits}). Both models gave an upper limit of X-ray luminosity of $1.3 \times 10^{30}~{\rm erg\ s^{-1}}$ (0.3$-$8 keV). 

\begin{table*}
    \centering
    \begin{tabular}{lcccccc}
         \hline
         MSP & {Spectral Model}$^\text{a}$ & ${R_\text{eff}}^\text{b}$ & $T_\text{eff}$ & Photon Index & {Reduced Stat}$^\text{c}$/Q-value & $F_X$ (0.3$-$8 keV) \\
          &  & (km) & ($10^6$ K) &  &  & ($10^{-15}$ erg cm$^{-2}$ s$^{-1}$) \\
         \hline
         A & BB & ${0.3^{+0.1}_{-0.1}}$ & $[1.6]^{\rm d}$ & $-$ & 0.09/0.77 & $0.2_{-0.1}^{+0.2}$ \\
           & PL & $-$ & $-$ & $[2.0]^{\rm d}$ & 0.31/0.58 & $0.2^{+0.1}_{-0.1}$ \\
         B & BB & ${0.02_{-0.02}^{+\infty}}^{\rm e}$ & ${7.4_{-7.4}^{+\infty}}^{\rm e}$ & $-$ & 1.55/0.17 & ${1.1_{-1.1}^{+\infty}}^{\rm e}$ \\
           & PL & $-$ & $-$ & $1.8 \pm 0.7$ & 1.11/0.35 & $1.4^{+0.7}_{-0.7}$ \\
         C & BB & ${0.4^{+0.3}_{-0.4}}^\text{f}$ & $1.5^{+0.4}_{-0.4}$ & $-$ & 0.23/0.88 & $0.6^{+0.2}_{-0.1}$ \\
           & PL & $-$ & $-$ & $3.9 \pm 1.2$ & 0.29/0.83 & $0.9^{+0.6}_{-0.6}$ \\
         D & BB & ${0.4^{+0.3}_{-0.4}}^\text{f}$ & $1.6^{+0.4}_{-0.4}$ & $-$ & 1.42/0.21 & $0.9^{+0.2}_{-0.1}$ \\
           & PL & $-$ & $-$ & $3.7 \pm 0.9$ & 2.05/0.07 & 1.3$^{+0.7}_{-0.7}$ \\
         E & BB & {0.07$^{+0.04}_{-0.07}$}$^{\rm f}$ & $3.8^{+1.0}_{-1.0}$ & $-$ & 0.84/0.58 & $1.2_{-0.9}^{+1.9}$ \\
           & PL & $-$ & $-$ & $2.2 \pm 0.6$ & 0.75/0.67 & $1.9^{+0.7}_{-0.7}$ \\
         F & BB & ${0.4^{+0.2}_{-0.4}}^\text{f}$ & $1.7_{-0.3}^{+0.3}$ & $-$ & 0.94/0.48 & $1.2^{+0.2}_{-0.1}$ \\
           & PL & $-$ & $-$ & $3.7 \pm 0.7$ & 1.04/0.40 & 2.0$^{+0.8}_{-0.8}$ \\
         \hline
    \end{tabular}
    \caption{Spectral fits for the M13 MSPs. 
    a) PL = power-law, BB = blackbody. The hydrogen column density was fixed to $N_\text{H} = 1.7 \times 10^{20}$ cm$^{-2}$. All uncertainties are 1 $\sigma$. 
    b) $R_\text{eff}$ calculated assuming a distance of 7.1 kpc. 
    c) Reduced statistic, calculated by the fit statistic divided by the degrees of freedom. 
    d) Obtained by averaging the best spectral fits of other MSPs. See section~\ref{subsec:spectral fits} for more details.
    e) Bounds unavailable.
    f) Model reached lower bound.}
    \label{tab:fits}
\end{table*}

\subsection{Optical/UV photometry}
We use the {\sc dolphot} software (version 2.0) to generate photometry catalogues for the WFC3 and ACS images. {\sc dolphot} is a photometry package based on {\sc HSTphot} \citep{Dolphin00} which provides pipelines to perform aperture and PSF photometry on individual FLC images. We first choose the $\U$ and $\I$ drizzle-combined images\footnote{Note that the drizzle-combined images used here are in their native pixel scales.} as the reference frames for WFC3 and ACS, on which {\sc dolphot} runs a detection algorithm to find stars. The reference images also provide master coordinates ($x, y$) to which stars on individual FLC frames are transformed. In the next step, we mask the flagged bad pixels in all FLC images using the {\tt wfc3mask} and {\tt acsmask} tools. These tools also multiply the FLC images and the pixel areas, converting the pixel units to electrons. The final photometry routine runs on separate CCD chips, which are extracted from the FLC images by the {\tt splitgroups} tool. These chip-specific images are also needed for the {\tt calcsky} tool to create corresponding sky images. With all these preparations, we finally run the {\tt dolphot} routine for WFC3 and ACS, adapting a photometry aperture ($\mathtt{img\_RAper}$, in pixels) of $8$, and a PSF radius ($\mathtt{img\_RPSF}$, in pixels) of $15$, while defining a sky annulus ($\mathtt{img\_RSky}$, in pixels) with inner and outer radii of $9$ and $14$, respectively, for PSF photometry. The final magnitudes are calibrated to the VEGMAG system using updated photometry zeropoints for ACS\footnote{\url{https://acszeropoints.stsci.edu/}} \citep{Sirianni05} and WFC3\footnote{\url{https://www.stsci.edu/hst/instrumentation/wfc3/data-analysis/photometric-calibration/uvis-photometric-calibration}}.

\begin{figure}
    \centering
    \includegraphics[width=\columnwidth]{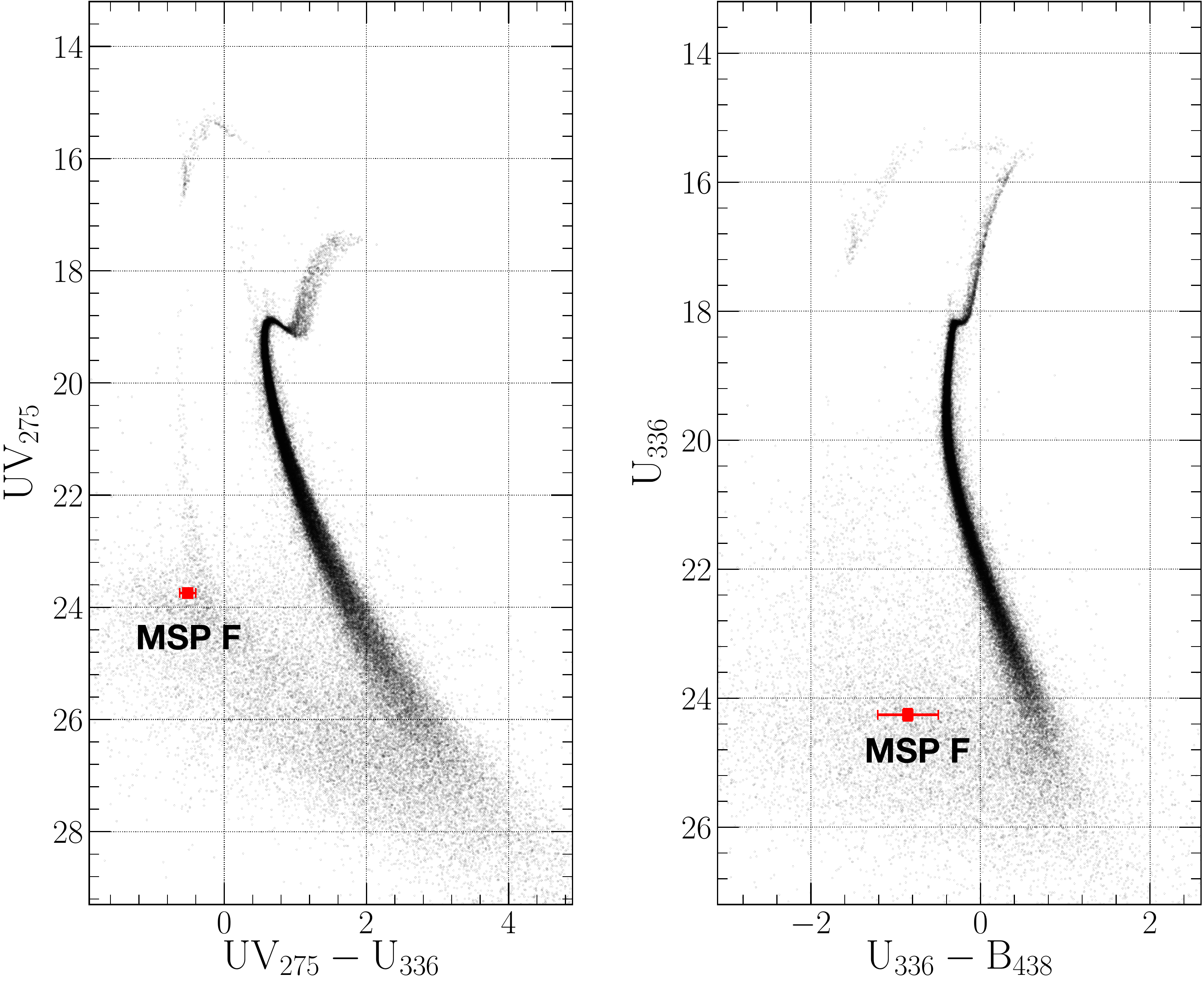}
    \caption{$\UV-\U$ and $\U-\B$ CMDs of M13. The red square marks the location of the counterpart to MSP F, which is bluer than the main sequence, consistent with the white dwarf cooling sequence.}
    \label{fig:wfc3_cmds}
\end{figure}

\begin{figure}
    \centering
    \includegraphics[width=\columnwidth]{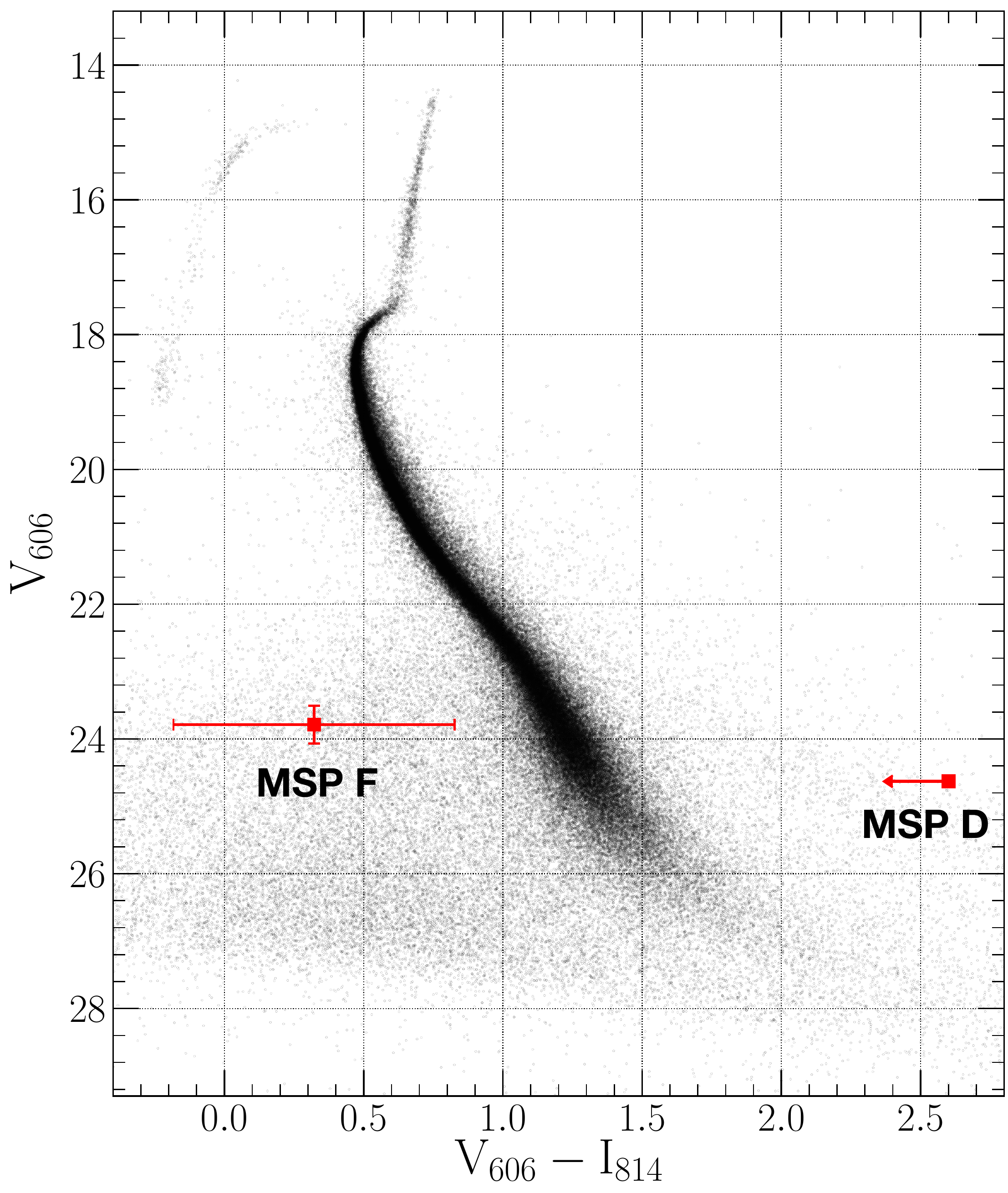}
    \caption{$\V-\I$ CMD of M13. Red squares mark the counterparts to MSP D and MSP F. The former was not detected in $\I$, so only a $3~\sigma$ upper limit is put on its $\V-\I$ colour.}
    \label{fig:acs_cmd}
\end{figure}

We keep stars with S/N>5 and rule out non-star objects, leaving cleaned catalogues to make colour-magnitude diagrams (CMDs; Figure \ref{fig:wfc3_cmds}, \ref{fig:acs_cmd}). These catalogues are used to compare the potential counterpart's photometry with the bulk of stars of the cluster.

\subsection{Optical/UV counterparts}
Since the radio timing solution provides much more accurate localisation ($\sim\mathrm{mas}$; \citealt{Wang20}) than the {\it Chandra} imaging (error radius $\sim 0.5\arcsec$), we expect potential counterparts in the vicinity of the corresponding radio position. We therefore search around the radio timing positions for optical counterparts in the drizzle-combined images; this leads to the discovery of optical/UV counterparts to two of the 6 MSPs: MSP D and F, both of which are very close ($\approx 0.02\arcsec$) to the radio positions. We report their photometric properties in the following paragraphs.

\subsubsection{MSP D}
The counterpart to MSP D is a faint star $15~\mathrm{mas}$ north from the radio timing position (Figure \ref{fig:finding_charts}). 
Although the counterpart is visible by visual inspection in the $\V$-band image (Figure \ref{fig:finding_charts}), it was not measured by {\sc dolphot}. We hereby make a rough estimate of its magnitude by performing aperture photometry, using an aperture of $0.1\arcsec$ to enclose most of the PSF of the star. Since the counterpart is in the vicinity of a very bright star, we also estimate background counts with the same aperture size in a nearby source-free region. The background-subtracted counts are then calibrated to {\sc dolphot} magnitude by $\V = -2.5\log_{10}(\text{net counts}) + 33.31$, giving $\V \approx 24.63$. We set a lower limit on the $\I$ band magnitude at $3$ times the local background counts, which gives $\I \gtrsim 22.03$. The upper limit on the $\V-\I$ colour is on the red side of the main sequence.

\subsubsection{MSP F}
\begin{figure*}
    \centering
    \includegraphics[width=\textwidth]{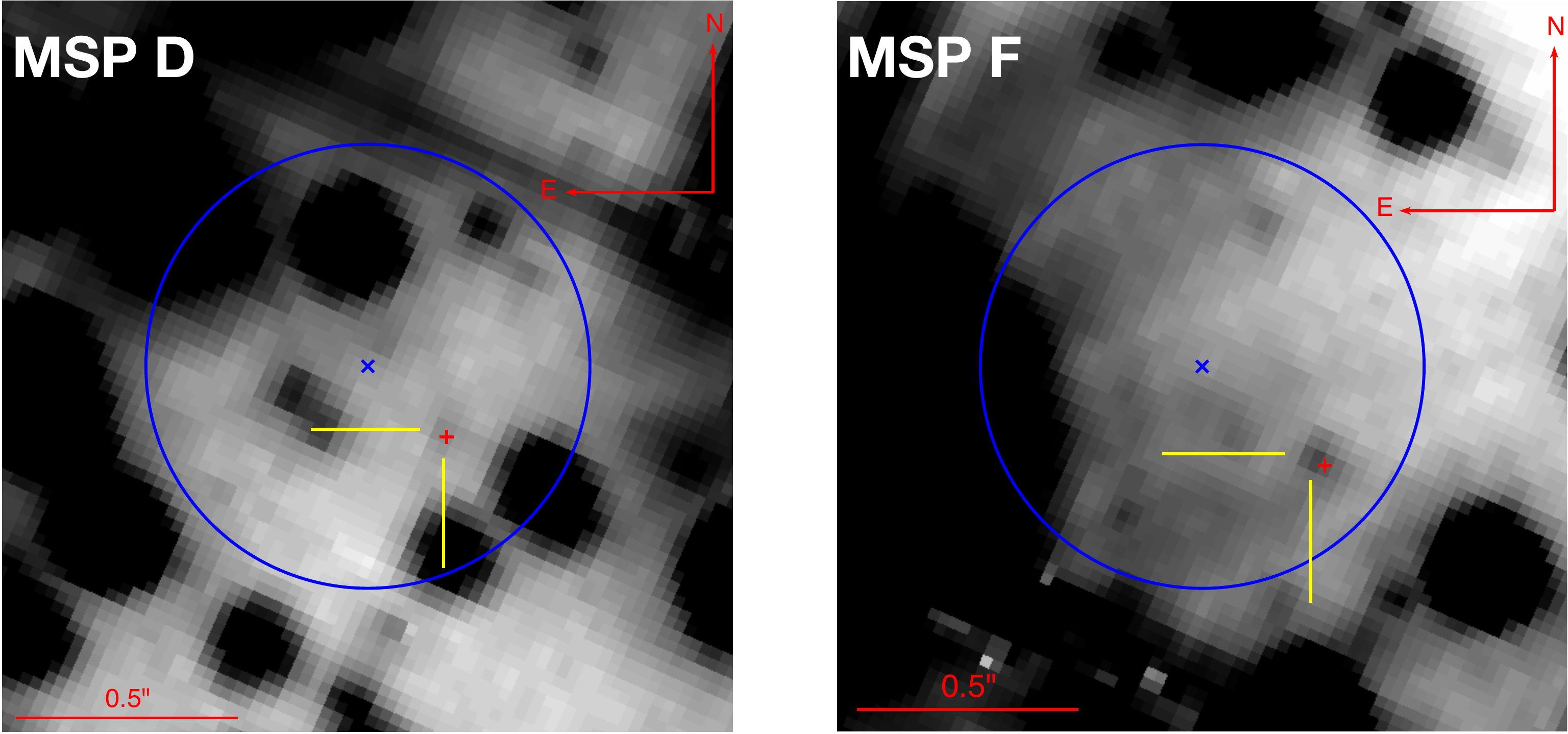}
    \caption{$\V$ finding charts for MSP D (left) and MSP F (right), showing a $1\farcs65 \times 1\farcs65$ square region centred on the X-ray position. North is up, and east is to the left. The blue cross in each chart indicates 
    the centre of the X-ray position error circle, while the blue circle shows the 95\% error region given the X-ray counts \citep{Hong05}. The red cross marks the radio timing position from \citet{Wang20}, and the optical counterparts are indicated with yellow crosshairs.}
    \label{fig:finding_charts}
\end{figure*}

We found a faint star $\approx 22~\mathrm{mas}$ east from the radio timing position (Figure \ref{fig:finding_charts}). 
The counterpart to MSP F appears to be bluer than the main sequence on all three CMDs (Figure \ref{fig:wfc3_cmds}, \ref{fig:acs_cmd}), and overlaps the white dwarf cooling sequence on the $\UV-\U$ CMD. This is expected in MSPs descended from binary evolution with a giant companion \citep[e.g.,][]{Stairs04}, wherein the NS exhausts the companion's envelope via continued mass accretion, resulting in a WD companion \citep[e.g.,][]{Sigurdsson03, Splaver05, Cadelano19}. 
After submission of this manuscript, we became aware of \citet{Cadelano20}, which independently identified the optical counterpart to the MSP M13 F, and used photometry in additional {\it HST} filters to  characterize the companion as a 0.23 $M_{\odot}$ He-core white dwarf. Our optical counterpart analysis is consistent with theirs, though not as constraining.

\subsubsection{Chance coincidence}
We estimate the number of chance coincidences ($N_c$) by dividing the cluster field into concentric annuli centered on the cluster, and calculating the probability of a coincidence within our search area. 
The radio position offers a much smaller search area than the X-ray position, so we use the radio error circle to assess the chance coincidence rate. The uncertainty in the radio position comes from the uncertainty in the radio timing position from \citet{Wang20}, which we estimate to be 3 mas at most (noting that the uncertainty in declination is missing from M13 F in \citealt{Wang20}'s Table 2), and from the uncertainty in {\it HST}'s absolute astrometry (10 mas, see \S 2.2), giving a total uncertainty in the radio position in the {\it HST} frame of $\sim$ 10 mas.

Based on the $\mathrm{UV_{275} - U_{336}}$ CMD, we count numbers of objects that align with the main sequence and the white dwarf sequence within each annulus. We did not apply proper-motion cleaning to the CMD, since we are comparing the estimated counts to actual observed numbers of sources in each search area. The counts are then divided by the annulus areas to give the $N_c$ per unit area, which is then multiplied by the search area of MSP D and MSP F (10 mas of radius) to give the $N_c$ per search area. We estimate $\approx$
0.001 
main sequence stars within the search region of MSP D, and $\approx$ $4\times10^{-6}$ WDs in the search region of MSP F. Both counterparts to these MSPs are therefore very unlikely to be chance coincidences.

\section{Discussion}
\label{sec: Discussion}

The majority of X-ray emission from most isolated MSPs is thermal radiation with blackbody-like spectra \citep[e.g.][]{Bogdanov06,Forestell14}, which is believed to result from polar cap heating from inverse Compton scattering \citep[e.g.][]{Harding02}. The observed spectrum of MSP C is a typical example of those of isolated MSPs. However, some relatively young  MSPs with high spin-down power, like B1821$-$24 in the globular cluster M28 \citep{Saito97,Bogdanov11}, 
produce pulsed non-thermal radiation generated by relativistic particles accelerated in the pulsar magnetosphere. 

Intriguingly, given the upper limit of X-ray luminosity of $1.3 \times 10^{30}~{\rm erg\ s^{-1}}$ (0.3$-$8 keV), MSP A is unusually faint, compared to other GC MSPs. We compare it to MSPs detected in Chandra X-ray observations that reached a sensitivity limit of $L_X<10^{30}~{\rm erg\ s^{-1}}$.
Only one MSP 
among 23 studied in X-rays 
in 47 Tuc \citep{Bogdanov06,Ridolfi16,Bhattacharya17}, namely PSR J$0024-7205$aa \citep{Bhattacharya17}, and one MSP among the  five MSPs in NGC 6752, namely PSR J1910$-$5959E \citep{Forestell14}, have X-ray luminosities slightly less than that upper limit of MSP A.
\citet{Bogdanov11} study 12 MSPs in M28, and report detections of 8 MSPs with $L_X$ between $1.0\times10^{30}$ for M28 D, and $10^{33}$ ${\rm erg\ s^{-1}}$ for M28 A, with three upper limits due to confusing, brighter sources. \footnote{\citet{Bogdanov11} report an upper limit $<1.0\times10^{30}$ ${\rm erg\ s^{-1}}$ for M28 I, but the position of M28 I used there proved to be incorrect, and the true $L_X$ of M28 I in its pulsar mode is $1-2\times10^{32}$ ${\rm erg\ s^{-1}}$ \citet{Papitto13,Linares14}. } 
Four other nearby globular clusters have deep X-ray and radio observations, with known MSPs: M22, NGC 6397, M71, and M4. Those four clusters each contain one, brighter, MSP  \citep{Grindlay01b,Bassa04,Elsner08, Amato19}. Thus, among 47 detectable MSPs in these eight  well-studied clusters, MSP A is one of the faintest four in X-rays (along with 47 Tuc aa, NGC 6752 E, and M28 D), and possibly the faintest. 
We cannot rule out either X-ray spectral model, by either Q-values or the fitted parameters, for MSP A. We prefer, however, the BB model, as MSP A is an isolated MSP. 
MSP A may have a relatively small spindown power, producing its relatively small X-ray luminosity. We cannot calculate its characteristic age or spindown power, due to its acceleration within the gravitational potential of M13, which produces an observed negative spin period derivative.
Alternatively, MSP A may appear to have a low X-ray luminosity due to an unfavorable inclination of its spin axis with respect to Earth,  with one or both hot spots on the far side from Earth (\citealt{Riley19} and \citealt{Miller19} showed that PSR J0030+0451 has both hot spots in a single hemisphere).

MSPs B, D, E, and F are in binary systems, with orbital periods ranging from 0.1 to 1.4 days \citep{Wang20}. The X-ray spectra of MSPs B and E are well described by a pure power-law model, indicating that the X-rays from them are predominantly non-thermal. Non-pulsed non-thermal emission is anticipated for MSP E, since it has been identified as an eclipsing black widow pulsar. In eclipsing spider pulsar systems, X-rays are thought to originate from intra-binary shocks, driven by the interaction of the relativistic pulsar wind with matter from its companion star \citep[e.g. MSPs J0023$-$7203J and J0024$-$7204W in 47 Tuc,][]{Bogdanov06}. Particularly, intra-binary shock emission with orbital modulation has been observed in many spider pulsar binaries \citep[e.g.][]{Bogdanov11b,Gentile14,AlNoori18}. 

The nature of the companion star of MSP B, which is in a 1.39-day orbit and has a minimum mass of 0.16$~\mathrm{M}_{\sun}$  \citep[assuming a pulsar mass of $1.4~\mathrm{M}_{\sun}$,][]{Wang20}, is not clear yet. 
The best-fit spectral model (a  power-law of photon index 1.8$\pm0.7$) requires either a magnetospheric origin (and thus a relatively high spindown energy loss rate to produce this), or an intrabinary shock origin, in which case (given the companion mass) MSP B would be a redback system.
We consider the 21 MSPs in the globular cluster 47 Tuc with X-ray spectra that can be individually fit \citep{Bogdanov06,Ridolfi16,Bhattacharya17}. The 17 systems that are not black widows and redbacks are all best fit by thermal blackbody spectra, while two of the three black widows (47 Tuc J,O,R), and the redback (47 Tuc W) all showed dominant power-law components to their spectra. Thus, it seems likely that MSP B is a redback system.
No eclipses were detected in the radio observations of MSP B \citep{Wang20}, which could be due to a relatively low inclination, as observed in the black widows 47 Tuc I and P \citep{Freire03}.
As intra-binary shocks are expected to emit X-rays in all directions (though not isotropically), the non-thermal X-rays may be detected at any inclination. 

The optical counterparts to MSPs D and F were found with {\it HST} observations  (Figure \ref{fig:finding_charts}). According to the position of the counterpart to MSP F on the CMDs of M13 (Figure~\ref{fig:wfc3_cmds}), as well as the minimum companion mass of $\sim 0.18~\mathrm{M}_\odot$, a white dwarf is the most likely companion star of MSP F. We also found a plausible optical counterpart to MSP D, which is only detected in the V band. We cannot identify the nature of MSP D's companion star definitively, due to its very faint magnitude leading to a large uncertainty of its location on the CMD (see Figure \ref{fig:acs_cmd}). 
The observed magnitudes are consistent with the rather broad expectations of MSP white dwarf companions  \citep{vanKerkwijk05}.

The X-ray spectrum of MSP D is well described by a pure blackbody model, implying that emission from the neutron star surface dominates, and that there is no intra-binary shock. Since all known redback binary systems show hard non-thermal X-ray emission \citep{Bogdanov18}, this indicates that MSP D is probably not in a redback binary. With a minimum companion mass of $\sim 0.18~\mathrm{M}_\odot$, we suggest that the companion star of MSP D is also a helium-core white dwarf. 


\section{Conclusions}
\label{sec: Conclusion}

In this report, we have presented X-ray and optical studies of the six MSPs in the globular cluster M13 by using archival {\it Chandra} and {\it HST} observations. Five of the six MSPs are firmly detected (MSPs B, C, D, E, and F) by {\it Chandra}, with X-ray luminosities $L_X \sim 3\times10^{30}-10^{31}$ erg s$^{-1}$ (0.3$-$8 keV), while MSP A is plausibly detected with an upper limit in X-ray luminosity of $1.3 \times 10^{30}~{\rm erg\ s^{-1}}$. The uncommonly X-ray-faint properties of MSP A may imply that one or both its hot spots are on the far side from Earth. 

The X-ray spectra of the six MSPs are well-described by either a single blackbody or a single power-law model. As expected, the spectra of two isolated MSPs, MSPs A and C, are well fitted by a pure blackbody model, indicating thermal X-ray emission from the surface of these two objects. The identified black widow binary system, MSP E \citep{Hessels07,Wang20}, emits principally non-thermal X-rays which are likely generated from intra-binary shock. Similarly, the X-ray emission from MSP B, a binary system with a companion of mass $\sim 0.2~{\rm M}_\odot$ \citep{Wang20}, is non-thermal as well. Based on its non-thermal spectral properties and companion mass, we suggest MSP B is a redback binary system. 

We searched for the optical counterparts to the four MSP binary systems in M13 using {\it HST} archival data in the vicinity of the respective radio timing positions, 
and discovered optical 
 counterparts to MSPs D and F.
 The position of the counterpart to MSP F on color-magnitude diagrams shows that the companion star is most likely a white dwarf. The counterpart to MSP D is faint and only observed in V band, resulting in a large uncertainty of its position on  color-magnitude diagrams. However, given MSP D's blackbody-like X-ray spectrum and companion mass of $\sim 0.2~{\rm M}_\odot$ \citep{Wang20}, 
 we argue that the counterpart to MSP D is also 
 likely to be a 
 helium-core white dwarf.
 To our knowledge, this is the first use of the X-ray properties of a radio millisecond pulsar to predict the nature of its companion star.

\section*{Acknowledgements}

COH acknowledges support from NSERC Discovery Grant RGPIN-2016-04602. This research has made use of data obtained from the Chandra Data Archive and the Chandra Source Catalog, and software provided by the Chandra X-ray Center (CXC) in the application packages CIAO and Sherpa. This research has made use of NASA’s Astrophysics Data System. This research is based on observations made with the NASA/ESA Hubble Space Telescope obtained from the Space Telescope Science Institute, which is operated by the Association of Universities for Research in Astronomy, Inc., under NASA contract NAS 5–26555. 

\section*{Data availability}

The {\it Chandra} data used in this paper are available in the Chandra Data Archive (\url{https://cxc.harvard.edu/cda/}) by searching the Observation ID listed in Table~\ref{tab:obs} in the Search and Retrieval interface, ChaSeR (\url{https://cda.harvard.edu/chaser/}). The {\it HST} data used in this work can be retrieved from the Mikulski Archive for Space Telescope (MAST) Portal (\url{https://mast.stsci.edu/portal/Mashup/Clients/Mast/Portal.html}) by searching the proposal IDs listed in Table~\ref{tab:hst_obs}.




\bibliographystyle{mnras}
\bibliography{example} 








\bsp	
\label{lastpage}
\end{document}